# Proposal for a Single-Molecule Field-Effect Transistor for Phonons

Marcos G. Menezes[1], Aldilene Saraiva-Souza[2], Jordan Del Nero[1,2] and Rodrigo B. Capaz[1]

[1]*Instituto de Física, Universidade Federal do Rio de Janeiro, Caixa Postal 68528, Rio de Janeiro, RJ 21941-972, Brazil*

[2]*Departamento de Física, Universidade Federal do Pará, Belém, PA 66075-110, Brazil*

## Abstract

We propose a practical realization of a field-effect transistor for phonons. Our device is based on a single ionic polymeric molecule and it gives modulations as large as -25% in the thermal conductance for feasible temperatures and electric field magnitudes. Such effect can be achieved by reversibly switching the acoustic torsion mode into an optical mode through the coupling the applied electric field to the dipole moments of the monomers. This device can pave the way to the future development of phononics at the nano or molecular scale.

Manipulating phonons with the same degree of control we manipulate electrons has always been an elusive dream for physicists. Indeed, even though we have a fairly complete understanding of both phonons and electrons from a theoretical point-of-view, as far as applications are concerned there is a massive imbalance between electronics and phononics. The electronic transistor was invented in 1947 [1] and it has revolutionized our lives since them, but so far there are no experimental realizations of its phonon counterpart. It is true that such imbalance can be traced back to the different quantum statistics of phonons (bosons) and electrons (fermions) and therefore is unavoidable. However, the potential range of applications that could arise from taming the flow of thermal current is so broad that it justifies a strong pursue of this goal.

Nanoscale science and engineering offer some hope in realizing this dream. The development of sophisticated lithography techniques at the nanoscale made possible a landmark achievement in this field in the year 2000 [2]: The measurement of the quantum of thermal conductance, theoretically predicted two years earlier [3]. Also, the recent realization of a thermal rectifier using mass-asymmetric carbon nanotubes is a promising step in the context of building thermal devices [4]. However, the realization of an *active* device such as a transistor remains an open subject. Not only experiments towards this goal are inexistent, but even theoretical proposals for such devices remain, to some extent, somewhat formal and generic [5].

In this work, we propose a practical realization of a single-molecule phonon field-effect transistor based on controlling the thermal conductance due to phonons in an ionic polymer. Our proposal has the following very attractive features:

(1) The phonon conductance is controlled by a back-gate electric field. This is highly desirable, since electrical operation allows much faster switching times and more precise control than mechanical switching.

(2) It works at the nano or molecular scale. This is important because, as we mentioned, the degree of control and measurement sensibility of the thermal energy flow is greater at this scale. Besides, the proposal can be implemented with the existing technology used in the field molecular electronics [6].

(3) It operates at achievable temperatures and electric field magnitudes.

Our device, schematically shown in Fig. 1, works in the ballistic regime [7]. In this regime, the heat current $\dot{Q}$ between hot and cold reservoirs with temperatures $T_H$ and $T_C$ respectively is given by the Landauer formula:

$$\dot{Q} = \sum_{\alpha} \int \frac{dk}{2\pi} \hbar\omega_\alpha(k) v_\alpha(k) \left[ n_{\alpha,k}(T_H) - n_{\alpha,k}(T_C) \right] , \tag{1}$$

where $\hbar\omega_\alpha(k)$ and $v_\alpha(k)$ are phonon energy and velocity for the $\alpha$ phonon branch with wavevector $k$ and $n_{\alpha,k}(T) = \left[\exp\left(\hbar\omega_\alpha(k)/k_B T\right) - 1\right]^{-1}$ is the phonon occupation number. After some algebra [3], the thermal conductance $\kappa$ in the linear response regime is given by

$$\kappa = \dot{Q}/\Delta T = \frac{k_B^2 \pi^2}{3h} T N_a + \frac{k_B^2}{h} T \sum_{\alpha'} \left[ \frac{\pi^2}{3} + f(x_0) + \frac{x_0^2 e^{x_0}}{e^{x_0} - 1} \right] , \tag{2}$$

where $\Delta T = T_H - T_C$ , $T = \left(T_H + T_C\right)/2$ , $f(x) = 2\mathrm{dilog}(e^x)$ [8] and $x_0 = \hbar\omega_{\alpha'}(0)/k_B T$ . In Eq. (2), the phonon conductance is written as sum of two contributions. The first term is the contribution from the $N_a$ acoustic branches, which gives rise to the quantum of thermal conductance at low temperatures [2,3]. Notice that this term is universal and therefore it does not depend on the details of the phonon dispersion. The second term is the contribution from all $\alpha'$ optical modes, which depends only on the values of the cutoff frequencies $\omega_{\alpha'}(0)$ , i.e., the minimum frequency associated with the phonon branch $\alpha'$ (not necessarily at $k = 0$).

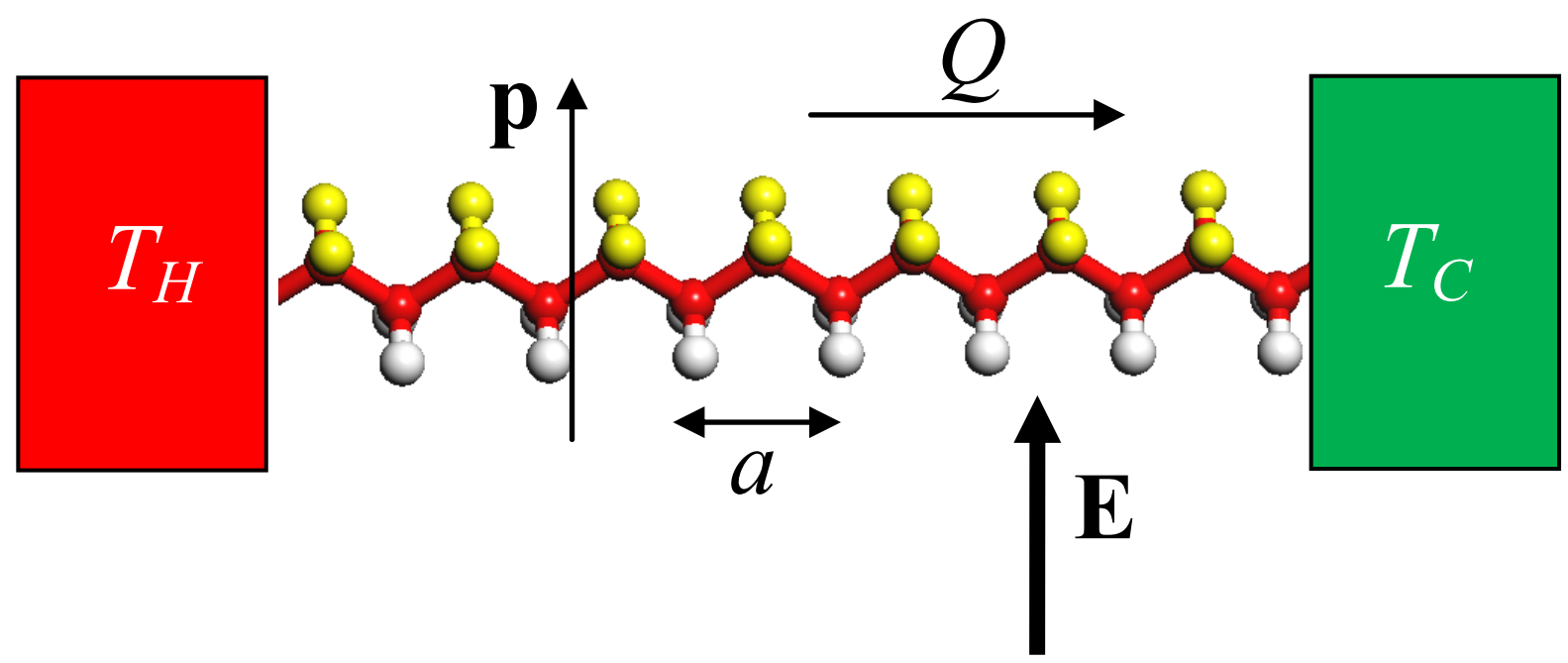

**Figure 1** – (Color online) Schematic representation of a single-molecule phonon field-effect transistor.

At a first glance, Eq. (2) offers very little hope of substantially modulating the phonon conductance by applying an electric field. After all, the contribution from the acoustic modes is universal, so any modification in the phonon dispersion caused by the electric field is not going to change the conductance. One could perhaps try to modify the contribution from the optical modes by modulating the cutoff frequencies $\omega_{\alpha'}(0)$. However, the contribution from optical modes to the conductance is exponentially small at low temperatures. Moreover, there is a more fundamental problem in that lattice vibrations do not usually couple strongly to external electric fields.

All these problems are solved by our appropriate choice of the active medium for the transistor: Highly ionic (or polar) polymers. Single polymeric molecules are an attractive system for exploiting thermal conduction in one dimension [11]. Ionic polymers have a permanent dipole moment **p** per monomer unit. In the absence of an external electric field, an infinite chain of such a polymer has 4 acoustic phonon branches, corresponding to 3 translations and one torsion branch. At $k \to 0$, the torsion branch is a free rotation of the chain, which costs zero energy. However, if an external electric field perpendicular to the chain axis is applied through a back gate (Fig. 1), the coupling between the field and the dipole moments breaks the full rotational symmetry of the chain, thus transforming the torsion branch from acoustic to optical. If the coupling is very effective, it will give rise to a substantial cutoff frequency for this branch. Then, at sufficiently low temperatures ($T << \omega_{\alpha'}(0)/k_B$), the phonons from this branch will be effectively blocked from participating in the heat flow and the thermal conductance will drop significantly. In fact, we anticipate a maximum modulation of -25% of the conductance using this effect, corresponding to blocking one out of four original acoustic modes. Therefore, the phonon field-effect conductance modulation in our device is caused by the *conversion of an acoustic to optical phonon branch by the application of an electric field*.

Let us now analyze in greater detail this prediction. Suppose each monomer unit *i* has a dipole moment $\mathbf{p}_i$ perpendicular to the chain and a moment of inertia *I* with respect to the chain axis. Then, the total energy associated with the torsion mode is given by

$$U = \frac{1}{2} I \sum_i \dot{\theta}_i^2 - \sum_i \vec{p}_i \cdot \vec{E} + \sum_i V(|\theta_i - \theta_{i+1}|) \,, \tag{3}$$

where the first term is the kinetic energy, the second term is the interaction energy between monomer dipoles and the external electric field $E$ and the $V$ is the torsion potential between first-neighbor monomers. Assuming that the ground-state configuration corresponds to all dipoles oriented in the same direction of the electric field ($\theta_i = 0$) [13], for small displacements from equilibrium the energy expression (3) can be written as

$$U = \sum_i \frac{1}{2} I \dot{\theta}_i^2 - pE \sum_i \left(1 - \frac{\theta_i^2}{2}\right) + \frac{1}{2} k \sum_i (\theta_i - \theta_{i+1})^2 \quad , \tag{4}$$

where $k$ is the angular restoring force constant in the harmonic approximation. This is a system of coupled 1D harmonic oscillators in an external field, and the related equation of motion can be easily solved with the normal mode transformation $\theta_i = Ae^{i(qx_i - \omega t)}$, giving the dispersion relation for torsion modes:

$$\omega(q)=\sqrt{\frac{pE+2k\left[1-\cos(qa)\right]}{I}}\,, \tag{5}$$

where $a$ is the polymer lattice constant. Eq. (5) shows the features we have anticipated: For zero field, the torsion mode is acoustic, with phonon velocity $v=\sqrt{\frac{ka^2}{I}}$. For non-zero field, the mode is optical, with a cutoff frequency at $q$ = 0 given by $\omega(0)=\sqrt{\frac{pE}{I}}$. This result, together with Eq. (2), gives us a recipe to maximize the thermal conductance modulation: One should search for polymers with the largest $p/I$ ratio, i.e., combining a large dipole moment with a small moment-of-inertia per unit.

We perform such a search by doing first-principles calculations within the density functional theory [14,15] and pseudopotential method. Exchange and correlation are treated within the local density approximation (LDA) [16]. We use norm-conserving Troullier-Martins pseudopotentials [17] and a numerical DZP (double-zeta with polarization) localized basis set. Calculations are performed using the SIESTA code [18]. Periodic boundary conditions are applied along the chain axis.

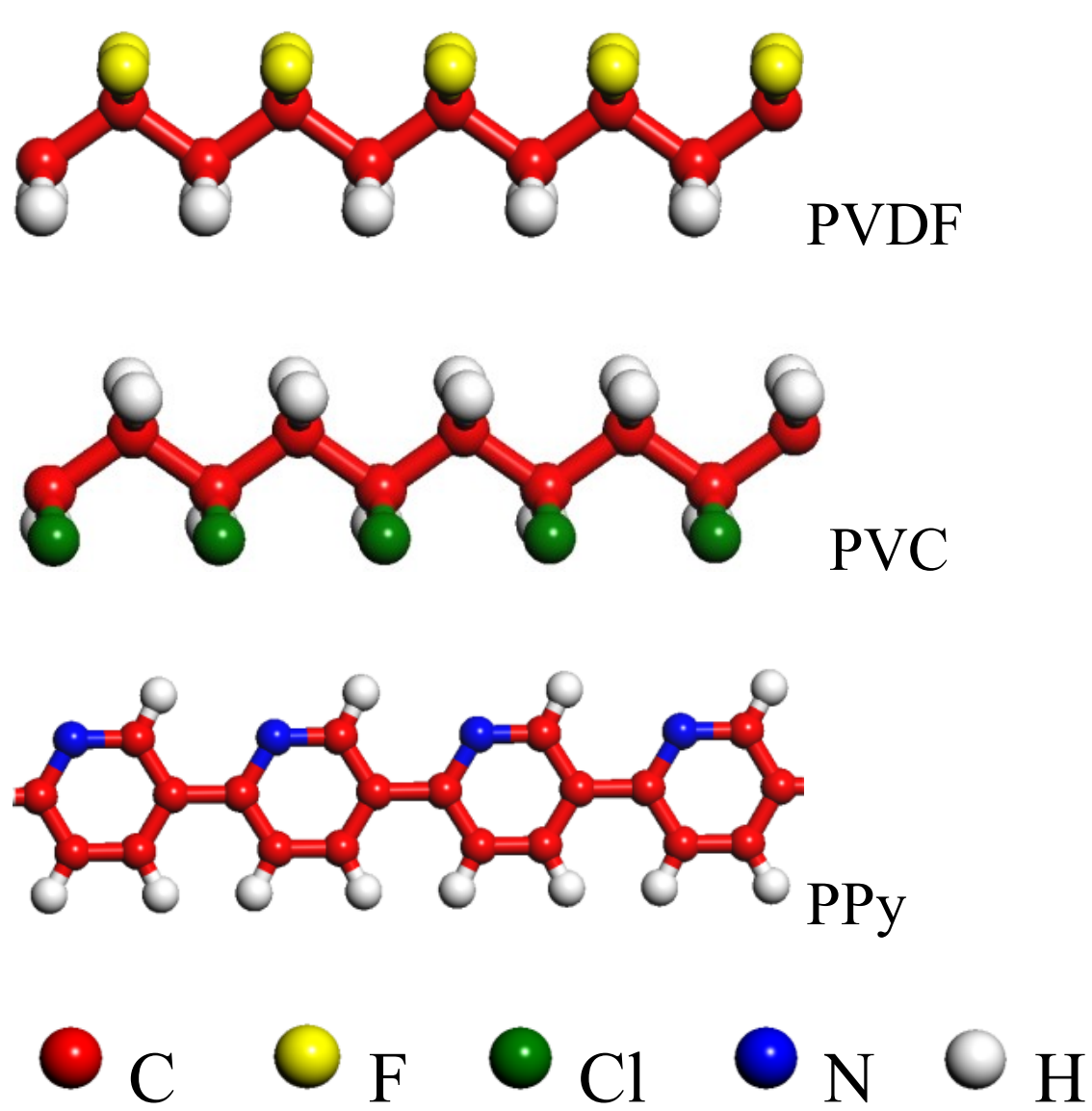

**Figure 2:** (Color online) Chemical structure of the three polymers investigated in this work.

We test three standard ionic polymers: Poly(vinylidene fluoride) – PVDF ($[C_2H_2F_2]_n$), polyvinyl chloride - PVC ($[C_2H_3Cl]_n$) and polypyridine – PPy ($[C_5NH_3]_n$), shown in Fig. 2. Table I shows the calculated dipole moments and moments-of inertia for all three polymers [19], together with the resulting values for the cutoff frequencies $\omega(0)$ for an electric field of 10 MV/cm (a typical field for nanoscale field-effect devices [20,21]). One sees that, among the three polymers investigated, PVDF is the one with largest value of $\omega(0)$. We express $\omega(0)$ in Kelvin to provide an estimate for the typical temperature for which our device will effectively block the thermal conductance associated with the torsion mode. From Table I, one sees that the device can work at liquid helium temperatures. However, even at somewhat larger temperatures, a substantial variation of thermal conductance can be achieved. This is shown in Fig. 3(a), where we plot the percent variation in thermal conductance (from Eq. (2)) as a function of temperature for PVDF at a 10MV/cm electric field. Notice that variations as large as -10% in conductance can be achieved for operating temperatures of roughly 12 K. In Fig. 3(b), we show, for a fixed temperature of $T$ = 4.2 K (liquid helium), the conductance modulation as a function of electric field for PVDF. Notice the sharp decrease of thermal conductance upon increasing the field.

| **Polymer** | ***p* (debye)** | ***I* ($10^{-45}$ kg.m$^2$)** | $\frac{\hbar\omega(0)}{k_B}$ **(K)** |
|---|---|---|---|
| PVDF | 1.55 | 1.32 | 4.8 |
| PVC | 1.25 | 1.36 | 4.2 |
| PPy | 1.43 | 1.46 | 4.3 |

**Table I:** Calculated dipole moment, moment-of-inertia and $\omega(0)$ (in Kelvin) for PVDF, PVC and PPy.

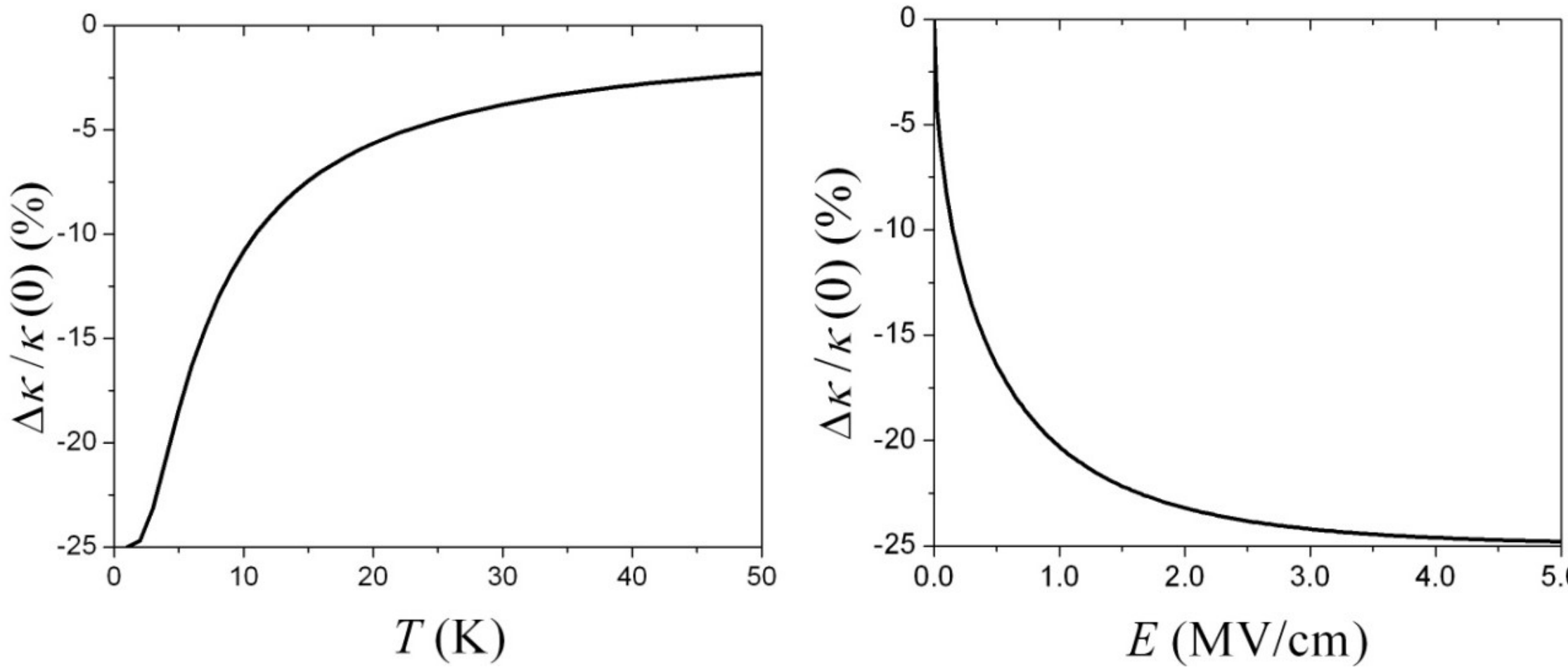

**Figure 3:** Percent variation of the thermal conductance of PVDF as a function of: (a) Temperature, for an applied electric field of 10 MV/cm; (b) Electric field magnitude, for a temperature of 4.2 K.

Finally, we should comment that, although, for simplicity, we have based our calculations on infinite chains, it is clear that the same effect of thermal conductance modulation must occur also in long (but finite) chains with attached ends to the thermal reservoirs. In this case, strictly speaking, the torsion mode at zero field will not be acoustic but it will have a small cutoff frequency, since the chain cannot rotate freely. However, for sufficient long chains, this cutoff frequency will be sufficiently small so that thermal conductance could be still modulated by an electric field [22]. In fact, based on the same arguments, we expect that some modulation of the thermal conductance can even occur for small ionic molecules in a break-junction geometry [22] or even in highly-oriented bulk polymers. Also, in our calculations, perfect phonon transmission is assumed between the reservoirs and the molecule for all phonon modes. If the transmission coefficient $T(k)$ for the phonon mode $k$ is smaller than one, the contribution from this mode to the thermal conductance will be scaled down by $T(k)$, as it can be seen from Eq. (1) of Ref. [3]. We cannot anticipate if including these effects will contribute to enhance or decrease the conductance modulation: This will depend on the relative values of $T(k)$ for all four acoustic branches, as well as their dependence on the electric field. We leave these issues open for future works.

In conclusion, we propose a realization of an electric field-effect transistor for phonons based on a single ionic polymeric molecule, in which modulations as large as -25% in the thermal conductance can be achieved at feasible temperatures and electric field magnitudes. The basic working principle of our device is the reversible switching of the acoustic torsion mode into an optical mode through the coupling of the applied electric field to the dipole moments of the monomers. Application of this effect can possibly lead to the first practical realization of a phonon field-effect transistor, which could pave the way to the future development of phononics.

**Acknowledgements:** We acknowledge financial support from Brazilian agencies CNPq, CAPES, FAPERJ, FAPESPA and INCT/Nanomateriais de Carbono.